\documentclass[%
 superscriptaddress,
 twocolumn,
 amsmath,amssymb,
 aps,superscriptaddress,
 prl,longbibliography
]{revtex4-2}

\usepackage[colorlinks=true,allcolors=blue]{hyperref} 
\usepackage{graphicx}
\usepackage{dcolumn}
\usepackage{bm}
\usepackage{physics} 
\newcolumntype{d}[1]{D{.}{.}{#1}} 
\usepackage{xcolor}
\usepackage{soul}
\usepackage{amsthm}

\usepackage{orcidlink}

\begin{document}


\title{Weak decays in superheavy nuclei}

\author{A. Ravli\'c\,\orcidlink{0000-0001-9639-5382}}
\email[]{ravlic@frib.msu.edu}
\affiliation{Facility for Rare Isotope Beams, Michigan State University, East Lansing, Michigan 48824, USA}

\author{W. Nazarewicz\,\orcidlink{0000-0002-8084-7425}}
\email[]{witek@frib.msu.edu}
\affiliation{Facility for Rare Isotope Beams, Michigan State University, East Lansing, Michigan 48824, USA}
\affiliation{Department of Physics and Astronomy, Michigan State University, East Lansing, Michigan 48824, USA}

\date{\today}

\begin{abstract}
Superheavy nuclei  represent the heaviest  atoms and nuclides known at the limit of mass and charge. The observed superheavy nuclei are all proton-rich; they  decay primarily by emitting $\alpha$ particles and fission, with a possible small electron capture (EC) branch. Due to the huge atomic numbers and associated relativistic effects, EC-decays of superheavy systems are expected to differ from what is known in lighter nuclei.
In this paper, using the quantified relativistic nuclear density functional theory and the quasiparticle random-phase approximation with the interaction optimized to experimental $\beta^-$-decay half-lives and Gamow-Teller resonance  energies,
we study the EC/$\beta^\pm$-decays in $Z = 101-118$ nuclei. Both allowed ($1^+$) and first-forbidden ($0^-, 1^-$ and $2^-$) transitions are considered.
We show that the first-forbidden $1^-$ transitions dominate the decay rates in almost all studied nuclei. For proton-rich nuclei, EC dominates over $\beta^+$ decay. We identify 44 nuclei with EC/$\beta^+$ branching ratio larger than 5\%, indicating a  possible competition with $\alpha$-decay and spontaneous fission channels.
\end{abstract}

\maketitle

\textbf{\textit{Introduction}}
The superheavy nuclei (SHN) occupy the region   of the chart of the nuclides, characterized by large atomic numbers, $Z \geq 104$. They provide a unique laboratory where nuclear forces compete with large  electrostatic repulsion \cite{Giuliani2019a,Smits2024a}. Due to highly relativistic electron motion, the superheavy atoms present a major challenge for atomic physics and chemistry \cite{Smits2023a,Smits2024a,Jerabek2018a}. At present, the heaviest discovered element is oganesson (Og, $Z = 118$) \cite{Oganessian2006a,Oganessian2012a}, while experimental efforts for $Z = 119$ and $Z = 120$ are underway \cite{Oganessian2009,Hofmann2016,Khuyagbaatar2020b,Sakai2022,Gates2024}. The main decay modes of the superheavy nuclei are $\alpha$-decay and spontaneous fission. Although occurring on a longer timescale compared to the strong-interaction decays, the $\beta^+$ and electron capture (EC) decays, mediated by the weak force, have been suggested to be present in  proton-rich SHN \cite{Smits2023a,Smits2024a,Heenen2015a,Khuyagbaatar2019,Khuyagbaatar2020}. 
Indeed, since the EC rate scales as $Z^3$, it becomes amplified in the superheavy region.

For the neutron-rich SHN, no experimental data exist. However, their $\beta^-$-decay rates could be relevant for the nucleosynthesis of heavy elements  through the astrophysical $r$-process \cite{Giuliani2018a,Giuliani2019a,Holmbeck2023a,Holmbeck2023b}.

From the theoretical side, the study of SHN is  challenging because of massive  extrapolations required into the regions where there are virtually no experimental information. The leading frameworks to study the properties of SHN are the macroscopic-microscopic approach \cite{Baran2005a,Sobiczewski2007a,Moller1994a} and nuclear  density functional (DFT) theory \cite{Cwiok1996,Cwiok2005a,Agbemava2015a,Agbemava2019a,Prassa2012a}. In this paper, we employ the  relativistic (covariant) DFT framework to study EC/$\beta^\pm$-decays in SHN. The early speculations that super-fast ECs ($T_{1/2} < 1$ s) could occur in SHN  motivated several studies based on schematic interactions~\cite{Sarriguren2019a,Sarriguren2020a,Sarriguren2021a,Sarriguren2022a,Moller1997a}, which concluded that weak decays should have a non-negligible branching ratio only for $Z < 108$. However, those studies considered only allowed decays, omitting the contribution of the forbidden transitions. In this paper, we  study allowed and first-forbidden EC/$\beta^+$ and $\beta^-$
decays in $101 \le Z \le 118$ nuclei using the state-of-the-art relativistic DFT based on point-coupling interactions \cite{Niksic2008a} and  relativistic quasiparticle random-phase approximation (RQRPA) \cite{Ravlic2024a}.

\textbf{\textit{Theoretical framework}} The nuclear ground state is calculated within the axially-deformed relativistic Hartree-Bogoliubov  theory, with the point-coupling DD-PC1 functional \cite{Niksic2008a}, assuming time-reversal and reflection symmetry \cite{Niksic2014a}. Odd-$A$ and odd-odd nuclei are calculated by blocking the quasiparticle configurations within the equal-filling approximation \cite{Schunk2010a}. For each nucleus we start from 4 values of initial quadrupole deformation in the range $|\beta_2| < 0.4$, avoiding super-deformed configurations \cite{Agbemava2015a}, and perform 20 constrained  iterations, after which the constraint is lifted, and calculations converge to the self-consistent local minima. For odd-$A$ and odd-odd nuclei, starting from an even-even reference state, we select a set of 5 blocking candidates for proton and neutron states. The configuration which minimizes the total energy is taken as the global minimum. The excited states are obtained with the relativistic quasiparticle random-phase approximation (RQRPA) in the linear response formulation \cite{Ravlic2024a}, with extension based on the equal-filling approximation \cite{Ney2020a,Shafer2016a}. The RQRPA time-odd residual interaction parameters that are unconstrained at the ground state level of  DD-PC1 are determined through a $\chi^2$-minimization on experimental data consisting of 26 $\beta^-$-decay half-lives and 4 Gamow-Teller (GT) resonance centroids, selected across the chart of the nuclides. Following Ref. \cite{Li2024a}, these parameters are: the isoscalar pairing strength $V_0^{pp}$, the Landau-Migdal strength $g_0$, and  the axial-vector coupling $g_A$. More details about the calibration approach can be found in  Sec. I of the  Supplemental Material  \cite{Sup}. 

Optimization is accomplished with the iterative derivative-free optimization software
POUNDERS  \cite{SWCHAP14}. Calculations for EC/$\beta^+$-decay are performed in a stretched harmonic-oscillator basis with $N_{\rm osc} = 18$ shells, while for $\beta^-$-decay, involving neutron-rich nuclei, we use $N_{\rm osc} = 20$ shells. This basis provides good convergence for moderate deformations considered in this paper ($|\beta_2| < 0.4$), setting an upper error estimate on half-lives related to the basis truncation not exceeding 3\%.

The EC rate in the lowest-order (LO) approximation of the electron radial wave functions has the form \cite{Bambynek1977a,Behrens1970a,Behrens1971a}
\begin{equation}\label{eq:first_expression_EC}
\lambda_{EC} = \frac{\ln{2}}{K} \sum \limits_x \sum \limits_{i, f} n_x C_x^{(i,f)} f_x(W_0^{(i,f)}),
\end{equation}
with $K = 6144$ s, where summation is performed over electron orbitals $x$, characterized by orbital angular momentum $l_x$ and total angular momentum $j_x$,
$n_x$ is the relative occupation of electrons in a given orbital, $C_x$ is the shape factor containing nuclear matrix elements, and $f_x$ is the phase-space factor 
$f_x = \frac{\pi }{2} q_x^2 \beta_x^2$,
where $q_x = W_0^{(i,f)} + W_x$ is the outgoing neutrino energy in units of $m_e c^2$, $W_0^{(i,f)}= E_i - E_f$ being the end-point energy, and $W_x$ is the orbital electron binding energy in units of $m_e c^2$. The information on lepton wave functions is contained in the Coulomb amplitude
$\beta_x$. Summation over $x$ in principle includes all electron orbitals in a superheavy atom, however, we restrict our sum to $\kappa_x = \pm 1$, where $\kappa_x$ is the relativistic block number. The Coulomb amplitudes $\beta_x$ are calculated by solving the radial Dirac equation for electron in a field generated by a superheavy nucleus, assuming a homogeneous distribution of charge $Z$ within the nuclear radius. To this end, we employ the DIRAC solver \cite{Salvat1995a}. We note that details of the nuclear charge distribution are not expected to impact the results \cite{Ravlic2024b,Smits2023a}. The $\beta^\pm$-decay can also contribute to the total decay rate, however, unlike electron being in a bound state, it is embedded in the continuum. The corresponding shape-factor $C^{(i,f)}(W)$ has a similar form as the EC shape-factor $C_x^{(i,f)}$, but can be expressed in powers of electron energy $W$~\cite{Bambynek1977a, Behrens1971a, Marketin2016a}.
 The total rate for the EC/$\beta^+$-decay is determined as $\lambda = \lambda_{EC} + \lambda_{\beta^+}$, while for $\beta^-$-decay, $\lambda = \lambda_{\beta^-}$. The half-life is $T_{1/2} = \ln(2)/\lambda$. Within the RQRPA, the summation over initial and final states $i,f$, with energies $E_i, E_f$, is replaced with summation over RQRPA eigenvalues $\nu$, with energies $\Omega_\nu$ \cite{Ravlic2021b}. In case of the linear response RQRPA, the rate can be expressed by contour integration in the complex plane over a suitably chosen contour encircling all the poles within the EC/$\beta^\pm$ energy window \cite{Ravlic2021a, Hinohara2013a, Ney2020a}.  The circular contour is  discretized with a 30-point grid using Gauss-Laguerre quadrature. The integration over electron energy $W$  is performed by Lagrange interpolation of the integrand on a 20-point Chebyshev grid, as suggested in Ref. \cite{Ney2020a}. Details are given in Sec. II.B of Ref.~\cite{Sup}. Calculations are performed for both allowed ($1^+$) and first-forbidden ($0^-, 1^-, 2^-$) multipoles.

\begin{figure}[htb]
    \includegraphics[width=0.87\linewidth]{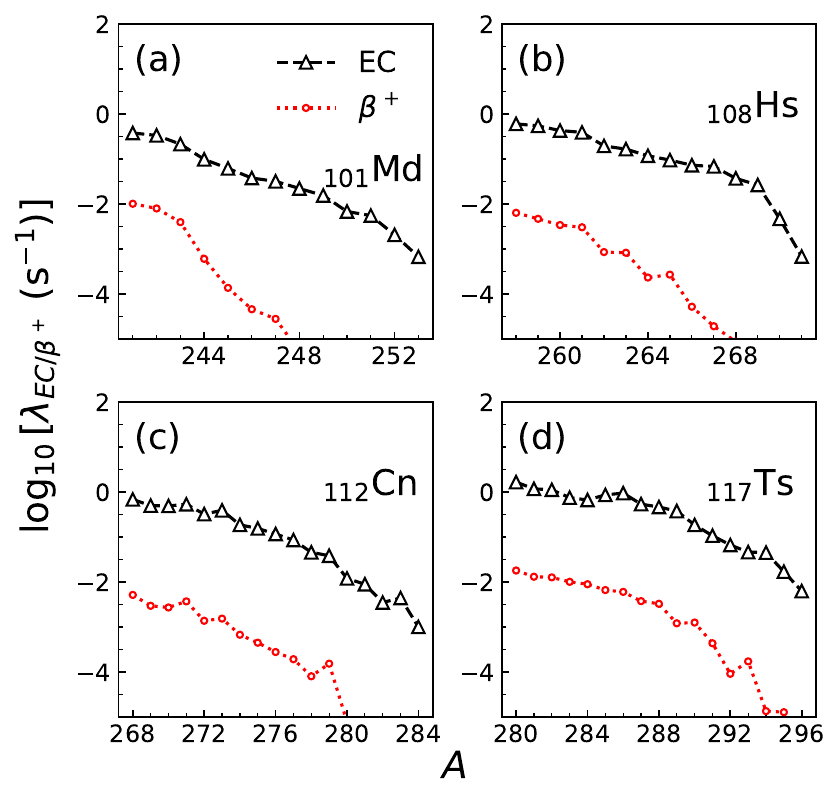}
    \caption{Comparison between the calculated EC and $\beta^+$-decay rates $\lambda_{EC/\beta^+}$ for (a)
    $_{101}$Md, (b) $_{108}$Hs, (c) $_{112}$Cn, and (d) $_{117}$Ts isotopic chains.}
    \label{fig:beta_vs_EC}
\end{figure}

\textbf{\textit{Results}} To study the interplay between  EC and $\beta^+$
decays in superheavy proton-rich region,  Fig.~\ref{fig:beta_vs_EC} shows the predicted rates, $\lambda_{EC/\beta^+}$, for the  isotopic chains of Md, Hs, Cn, and Ts.
In all cases, EC dominates the half-lives by more than an order of magnitude. At larger neutron numbers, the $\beta^+$-decay half-lives increase considerably faster than 
those due to  EC. This is an expected result since the EC rate  scales as $Z^3$ and becomes more important at large atomic numbers. On the other hand, for nuclei closer to the valley of beta stability, the $\beta^+$-decay phase-space factor decreases considerably with decreasing end-point energy $W_0$, since, for allowed transitions, it scales as $W_0^5$.

\begin{figure}[htb]
    \includegraphics[width=0.87\linewidth]{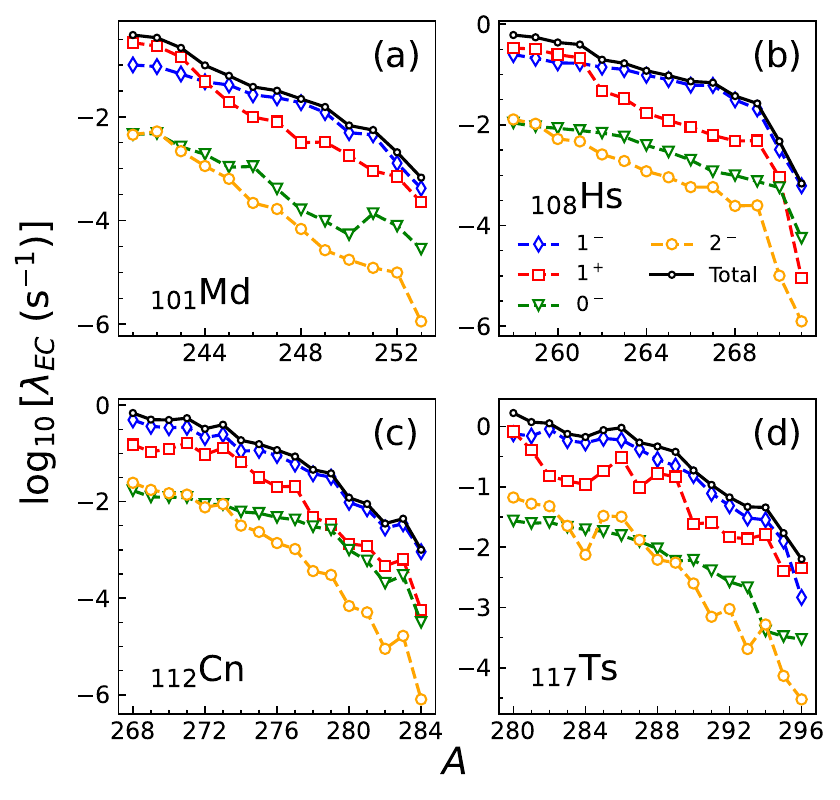}
    \caption{The decomposition of the total EC rate $\lambda_{EC}$ into allowed ($1^+$) and first-forbidden multipoles ($0^-, 1^-, 2^-$) for (a) ${}_{101}$Md, (b) ${}_{108}$Hs, (c) ${}_{112}$Cn,  and (d) ${}_{117}$Ts isotopic chains.}
    \label{fig:multipoles}
\end{figure}

To determine which multipole transitions contribute most to the decays, in Fig. \ref{fig:multipoles}, we show
contributions from allowed Gamow-Teller (GT) and first-forbidden (FF) transitions
to the total EC rate. We note that most calculations performed in the proton-rich SHE region up to date consider only allowed GT transitions, neglecting  FF contributions \cite{Sarriguren2019a,Sarriguren2020a,Sarriguren2021a,Sarriguren2022a,Moller1997a}. According to our calculations, the $1^-$ FF multipole tends to dominate the EC rate in almost all nuclei. In the case of proton-rich isotopes of Md and Hs, GT decay tends to compete with FF $1^-$ multipole, however, both are necessary to accurately determine the total EC decay rate. The contribution of $0^-$ and $2^-$ FF multipoles can be neglected, as they are predicted to have significantly lower rates. Contributions of different operators at the LO to the $1^-$ FF rate are studied in Sec. II.A of Ref.~\cite{Sup}.

\begin{figure}[ht!]
    \includegraphics[width=0.9\linewidth]{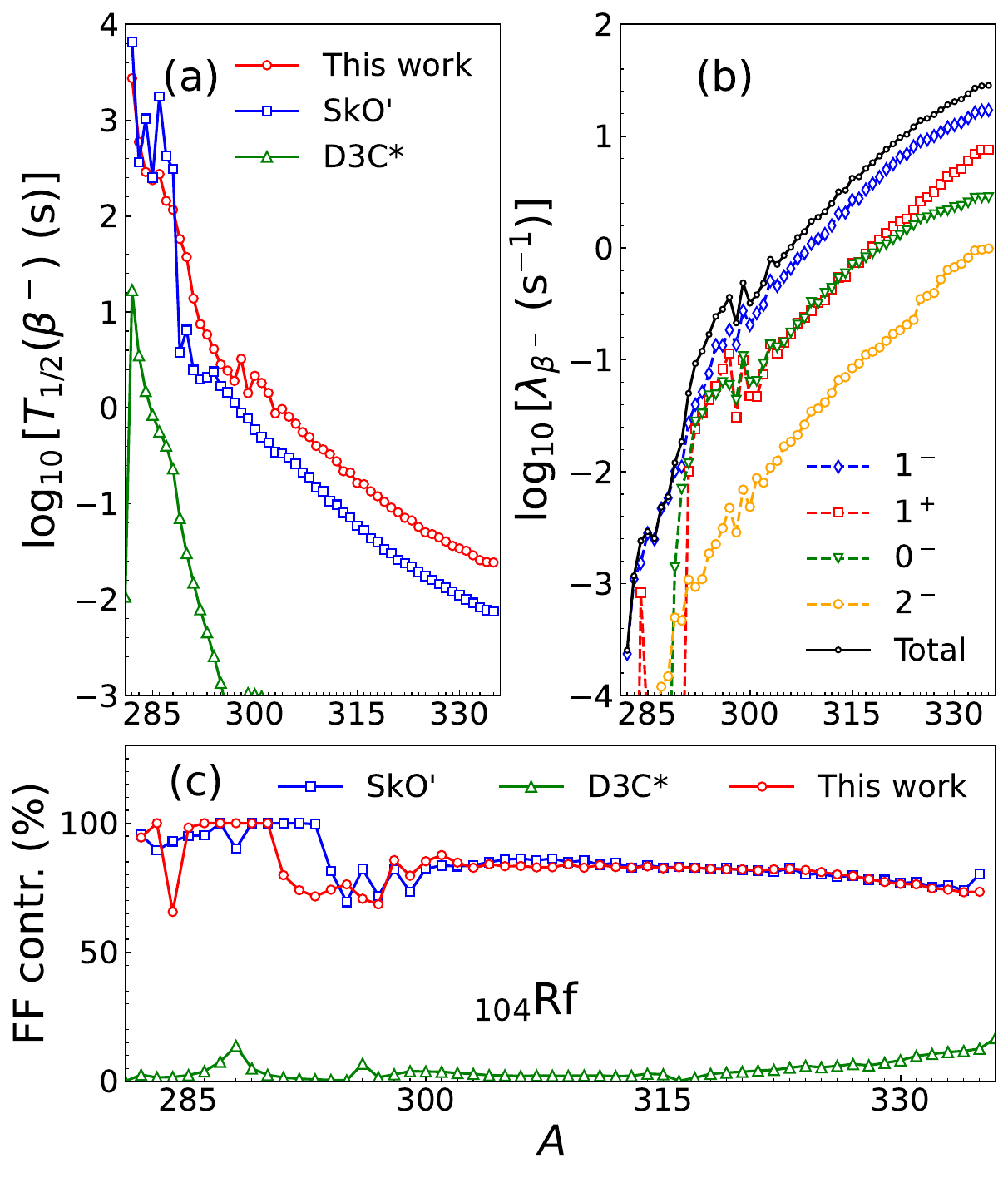}
    \caption{(a) $\beta^-$-decay half-lives for the Rf isotopic chain calculated in this paper (circles) compared to results of non-relativistic DFT calculations based on SkO' functional \cite{Ney2020a} (squares) and spherical relativistic D3C*  calculations  \cite{Marketin2016a} (triangles). (b) The partial contribution of allowed ($1^+$) and first-forbidden ($0^-,1^-,2^-$) transitions to the total decay rate $\lambda_{\beta^-}$. (c) The relative contribution of first-forbidden transitions (in \%) compared to Refs. \cite{Ney2020a,Marketin2016a}.}
    \label{fig:beta_m}
\end{figure}

Within the same formalism, we make predictions for $\beta^-$ decays in neutron-rich SHN. Since there exist no experimental data for $\beta^-$-decays in SHN, we compare our results with other theoretical predictions to ascertain systematic uncertainty. The results for the $_{104}$Rf isotopic chain  
are presented  in Fig. \ref{fig:beta_m} (the results for Db, Sg, Bh, and Hs chains are shown in Sec.~IV of Ref.~\cite{Sup}).
In panel (a) we compare our results with  non-relativistic SkO' deformed calculations of Ref. \cite{Ney2020a}, and relativistic D3C* spherical calculations of Ref. \cite{Marketin2016a}. We observe that two axially-deformed calculations yield fairly similar predictions, with differences within one order of the magnitude up to the neutron dripline. On the other hand, spherical calculations predict much shorter half-lives.
As in the case of EC/$\beta^+$-decay, the $1^-$ FF transitions dominate the $\beta^-$ decay rate, as can be seen in Fig. \ref{fig:beta_m}(b). By comparing the relative contribution of FF transitions to the total decay rate in Fig.\,\ref{fig:beta_m}(c), we observe that axially-deformed results from this paper and non-relativistic calculations in Ref. \cite{Ney2020a} predict very similar contribution of FF transitions for neutron rich nuclei. On the contrary, calculations from Ref. \cite{Marketin2016a} dramatically underestimate the relative importance of FF transitions.

\begin{figure*}[ht!]
    \centering
    \includegraphics[width=0.85\linewidth]{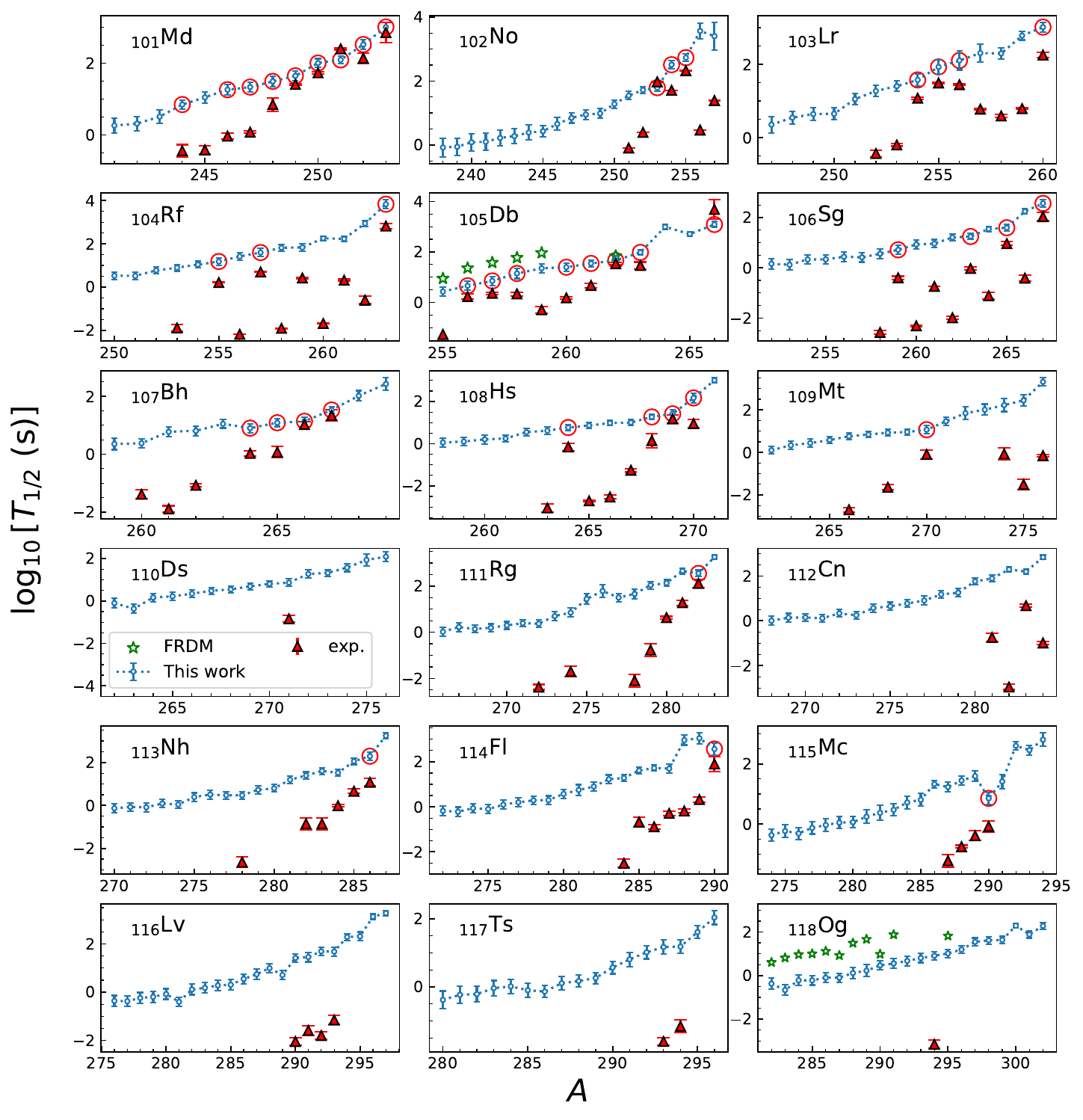}
    \caption{The EC/$\beta^+$-decay rates for nuclei between ${}_{101}$Md up to ${}_{118}$Og. The RQRPA calculations together with statistical uncertainties stemming from residual interaction (open  circles) are compared with the available experimental data (triangles) \cite{Kondev2021a}, and microscopic-macroscopic (FRDM) calculations (stars) \cite{Moller1997a}. Nuclei with EC branching ratio $\mathcal{R}$ predicted to be larger than 5\% with respect to measured half-lives are marked with red circles.}
    \label{fig:total_results}
\end{figure*}

In the following, we compare the calculated EC/$\beta^+$ decay half-lives with experimentally measured half-lives in the SHE region. Based on the comparison we would like to infer whether there exist superheavy nuclei for which the EC/$\beta^+$ decay has a non-negligible, or measurable  branching. To estimate the uncertainty of our predictions, we carry out the $\chi^2$-minimization of residual interaction parameters $V_0^{pp}, g_0$, and the axial-vector coupling $g_A$, and we use the resulting covariances to estimate theoretical errors, see Sec.~I Ref.~\cite{Sup} for details.

The predicted EC/$\beta^+$ decay half-lives for isotopic chains  $Z = 101-118$  are shown in Fig. \ref{fig:total_results}, together with experimental half-lives \cite{Kondev2021a}.
(The actual values are tabulated in Table S2 of Ref.~\cite{Sup}.) Starting close to the proton drip line, we have included those nuclei for which $T_{1/2} < 10^4$ s. We observe that for each isotopic chain, half-lives start around 1\,s, and  increase with $N$. No significant odd-even staggering is observed for odd-$A$ and odd-odd nuclei and isotopic dependence of half-lives seems to be fairly smooth. Especially interesting are nuclei for which EC has non-negligible contribution to the total decay rate. In fact, if we consider SHN for which the EC branching ratio $\mathcal{R}$, is larger than 5\%, then most candidates are found in odd-$Z$ chains up to $Z = 105$: nine in  mendelevium , four in lawrencium, and eight in dubnium. For even-$Z$ chains, three candidates are found in isotopic chains of nobelium and rutherfordium, and four in seaborgium and hassium. Out of a total of 44 candidates with EC branching larger than 5\%, only five are even-even, while all others are odd-$A$ and odd-odd. This is expected 
due to configuration differences  between parent and daughter nuclei in $\alpha$-decay that result in increasing  half-lives, and similar holds for spontaneous fission. The heaviest nucleus with non-negligible EC branching is predicted to be ${}^{290}$Mc. In addition, in Fig. \ref{fig:total_results}, we compare our results with the finite-range droplet model (FRDM) calculations from Ref. \cite{Moller1997a} for Db and Og. In general, those calculations tend to predict longer half-lives compared to ours, and especially so in Og, where the differences can be up to an order of magnitude. For instance, in Ref. \cite{Nelson2008a}, the EC branching for ${}^{256}$Db was measured to be around 0.3. Our calculations predict $0.37 \pm 0.17$, while the FRDM predicts 0.07. Further experiments searching for EC decays in SHN  would provide valuable  constraints on  calculations.

\textbf{\textit{Conclusions}}
In this paper, using the quantified relativistic EDF+QRPA model,  we investigated the weak decays of superheavy nuclei: EC/$\beta^+$-decays for proton-rich nuclei and $\beta^-$-decays in neutron-rich nuclei. In both cases, we show that the weak-decay rates are dominated by the first-forbidden transitions. For proton-rich superheavy nuclei, we show that weak decays are dominated by EC. We provide a list of 44  nuclei, primarily   with $Z \leq 108$, that could guide experimental efforts to unravel new decay chains of proton-rich SHN and study the phenomenon of EC at very large atomic numbers of   superheavy atoms.  For heavier isotopes, $\alpha$-decay and spontaneous fission are expected to dominate. As exemplified in our calculations, the fastest EC decays correspond to half-lives around  1\,s, while the fission and $\alpha$-decay lifetimes in this mass region are in a millisecond range. 

We wish to point out that the residual QRPA interaction used in our calculations has been calibrated to known GT decays and giant resonances; this allowed us to estimate related uncertainties on predictions. It is satisfying to see a good agreement between our predictions for $\beta^-$ lifetimes and DFT results of Ref.~\cite{Ney2020a}.
More detailed investigations, including different functionals and complete uncertainty quantification are left for future work. 

\textit{Acknowledgments --}
Discussions with Peter Schwerdtfeger and Sudhanva Lalit are gratefully acknowledged. This work was supported by the U.S. Department of Energy under Award Number DOE-DE-NA0004074 (NNSA, the Stewardship Science Academic Alliances program), and by the Office of Science, Office of Nuclear Physics under grants DE-SC0013365 and DE-SC0023175 (Office of Advanced Scientific Computing Research and Office of Nuclear Physics, Scientific Discovery through Advanced Computing). This work was also supported in part through computational resources and services provided by the Institute for Cyber-Enabled Research at Michigan State University.

%

\end{document}



\title{Supplemental Material: Weak decays in superheavy nuclei}

\author{A. Ravli\'c\,\orcidlink{0000-0001-9639-5382}}
\email[]{ravlic@frib.msu.edu}
\affiliation{Facility for Rare Isotope Beams, Michigan State University, East Lansing, Michigan 48824, USA}

\author{W. Nazarewicz\,\orcidlink{0000-0002-8084-7425}}
\email[]{witek@frib.msu.edu}
\affiliation{Facility for Rare Isotope Beams, Michigan State University, East Lansing, Michigan 48824, USA}
\affiliation{Department of Physics and Astronomy, Michigan State University, East Lansing, Michigan 48824, USA}

\date{\today}

\maketitle

This Supplemental Material contains more details on  calibration of the pnQRPA residual interaction parameters; details of weak transition rates calculations;  Table~\ref{betam-table} of 
EC/$\beta^+$ branching ratios; and results  for Db, Sg, Bh, and Hs $\beta^-$-decay chains.

\section{Optimizing the pnQRPA parameters}
The pnQRPA calculations involve residual interaction terms whose parameters are not constrained at the ground-state level of the relativistic Hartree-Bogoliubov (RHB) equation. In particular, in the particle-hole channel ($ph$) those include the isovector-pseudovector (TPV) term of the form \cite{Ravlic2021a,Ravlic2024a}:
\begin{align}
\begin{split}
V^{TPV}_{p n n^\prime p^\prime} &= g_0 \int d^3 \boldsymbol{r}_1 d^3 \boldsymbol{r}_2 \left[\bar{\Psi}_p(\boldsymbol{r}_1) \gamma_5^{(1)} \gamma_\mu^{(1)} \vec{\tau}^{(1)} \Psi_n(\boldsymbol{r}_1) \right] 
\left[\bar{\Psi}_{n^\prime}(\boldsymbol{r}_2) \gamma_5^{(2)} \gamma^{\mu (2)} \vec{\tau}^{(2)} \Psi_{p^\prime}(\boldsymbol{r}_2) \right]\delta(\boldsymbol{r}_1 - \boldsymbol{r}_2),
\end{split}
\end{align}
where $\Psi_{p(n)}(\boldsymbol{r})$ are proton(neutron) Dirac wavefunctions, $\boldsymbol{\tau}$ is the Pauli isospin matrix, and $\bar{\Psi} = \Psi^\dag \gamma_0$, $g_0$ is the so-called Landau-Migdal strength, usually determined by reproducing the experimental giant resonance energy centroids. In the particle-particle ($pp$) channel we assume the separable pairing form \cite{Tian2009a}:
\begin{equation}
V^\prime(\boldsymbol{r}_1, \boldsymbol{r}_2, \boldsymbol{r}_1^\prime, \boldsymbol{r}_2^\prime) = -f G\delta(\boldsymbol{R} - \boldsymbol{R}^\prime) P(r,z) P(r^\prime, z^\prime),
\end{equation}
where $\boldsymbol{R} = \frac{1}{2}(\boldsymbol{r}_1 + \boldsymbol{r}_2)$ is the center-of-mass and $\boldsymbol{r} = \boldsymbol{r}_1 - \boldsymbol{r}_2$ is the relative coordinate. The overall factor $f$ is defined as
\begin{equation}
    f = \left\{ \begin{array}{c}
         V_0^{pp}, \quad T = 0, S = 1  \\
         1, \quad T = 1, S = 0 
    \end{array} \right. ,
\end{equation}
where $V_0^{pp}$ is the isoscalar pairing strength.
The form factor $P(r,z)$ corresponds to the Gaussian function
\begin{equation}
P(r,z) = \frac{1}{(4 \pi a^2)^{3/2}} e^{- \frac{z^2 + r^2}{4a^2}}.
\end{equation}
The pairing strength $G$ and the range $a$ are adjusted to reproduce the pairing gap of the Gogny pairing force \cite{Tian2009a}.

In addition to $g_0$ and $V_0^{pp}$, we also include the strength of the axial-vector coupling $g_A$ in the set of unknown parameters. Following  Ref. \cite{Li2024a}, 
we carry out a $\chi^2$ minimization with respect to the parameters $\boldsymbol{x} \equiv  ( V_0^{pp}, g_A, g_0)$
to a dataset of 26 experimentally measured $\beta^-$-decay half-lives  in the range of $10^{-2}-10^3$\,s and 4 Gamow-Teller (GT) resonance centroid energies shown in Table \ref{tab:optimized_dataset}.

The penalty function for the optimization is defined as:
    \begin{equation}
\chi^2(\boldsymbol{x}) \sim \sum \limits_{k \in \beta} \left( \frac{\text{log}s_k(\boldsymbol{x}) - \text{log}d_k}{w_\beta} \right)^2 + \sum \limits_{k \in \text{GTR}} \left( \frac{s_k(\boldsymbol{x}) - d_k}{w_\text{GTR}} \right)^2, 
\end{equation}
where $s_k(\boldsymbol{x})$ is the half-life (or GT resonance energy) obtained with the pnQRPA for a set of parameters $\boldsymbol{x}$, $d_k$  is the experimental half-life (or resonance energy), and $w_\beta$, $w_{GT}$, are weights for half-life and GTR datasets, respectively. In this optimization, we use the values $w_\beta = 0.5$ and $w_{GTR} = 0.4$, which follow from approximating the distribution of weights and residuals for GT resonances as outlined in Ref. \cite{Dobaczewski2014a}.

\begin{table}
    \caption{Experimental $\beta^-$-decay half-lives (left), and GT resonance centroid energies (right) of nuclei used in the optimization of the parameters $(V_0^{pp},g_A,g_0)$.}\label{tab:optimized_dataset}
    \begin{tabular}{rr|rr}
    \hline
    \hline
    \\[-8pt]
         nucleus & $T^{exp.}_{1/2}$ (s)  & nucleus & $T^{exp.}_{1/2}$ (s)   \\
        \hline
            \\[-8pt]
         ${}^{58}$Ti & 0.058  & ${}^{60}$Cr & 0.49  \\
         ${}^{78}$Zn & 1.47  & ${}^{152}$Ce & 1.4  \\
          ${}^{48}$Ar & 0.416  & ${}^{96}$Kr & 0.080  \\
          ${}^{134}$Sn & 1.050  & ${}^{166}$Gd & 4.8  \\
          ${}^{100}$Zr & 7.1  & ${}^{108}$Mo & 1.09  \\
           ${}^{242}$U & 1008  & ${}^{142}$Xe & 1.23  \\
           ${}^{82}$Zn & 0.166  & ${}^{114}$Ru & 0.54  \\
           ${}^{156}$Nd & 5.26  & ${}^{236}$Th & 2250  \\
            ${}^{66}$Fe & 0.440  & ${}^{120}$Pd & 0.492  \\
          ${}^{162}$Sm & 2.4  & ${}^{88}$Se & 1.53  \\
           ${}^{102}$Sr & 0.069 & ${}^{228}$Rn & 65  \\
          ${}^{96}$Sr & 1.07  & ${}^{174}$Er & 192 \\
            ${}^{204}$Pt & 10.3  &  ${}^{72}$Ni & 1.57   \\
         \hline
    \end{tabular}
    \begin{tabular}{ccccc}
     \hline
     \hline     \\[-8pt]
       & ${}^{208}$Pb & ${}^{132}$Sn & ${}^{90}$Zr & ${}^{112}$Sn  \\
    \hline
       $E_{GT}$ (MeV)         & 19.2  & 13.2 & 15.5 & 16.8  \\
    \hline
    \end{tabular}
\end{table}

Optimization is accomplished with the 
iterative derivative-free
optimization software
POUNDERS  \cite{SWCHAP14}. The resulting optimal values of the parameters 
$\bar{\boldsymbol{x}}$ are:
\begin{equation}
  \bar{V}_0^{pp} =  1.254,  \quad \bar{g}_A =   1.216, \quad \bar{g}_0 =   0.622.
\end{equation}

In order to estimate standard deviations around optimal parameters we compute the Jacobian $\hat{J}$  using the finite-difference formula:
\begin{equation}
\hat{J}_{kl}(\bar{\boldsymbol{x}}) \approx \frac{\varepsilon_k(\bar{\boldsymbol{x}}+d\boldsymbol{e}_l) - \varepsilon_k(\bar{\boldsymbol{x}}-d\boldsymbol{e}_l)}{2d},
\end{equation}
where $\varepsilon_k$ is the weighted residual, defined as
\begin{equation}
    \varepsilon_k = 
    \left\{ 
    \begin{array}{c}
     (\log s_k(\boldsymbol{x}) - \log d_k )/w_\beta, \quad k \in \beta, \\
     (s_k(\boldsymbol{x}) -  d_k )/w_{\text{GTR}}, \quad k \in \text{GTR}, \\
    \end{array}
     \right.
\end{equation}
where  $d=10^{-3}$ and $\boldsymbol{e}_l$ is the unit vector pointing in the direction of model parameters.

\begin{figure}
    \includegraphics[scale = 0.7]{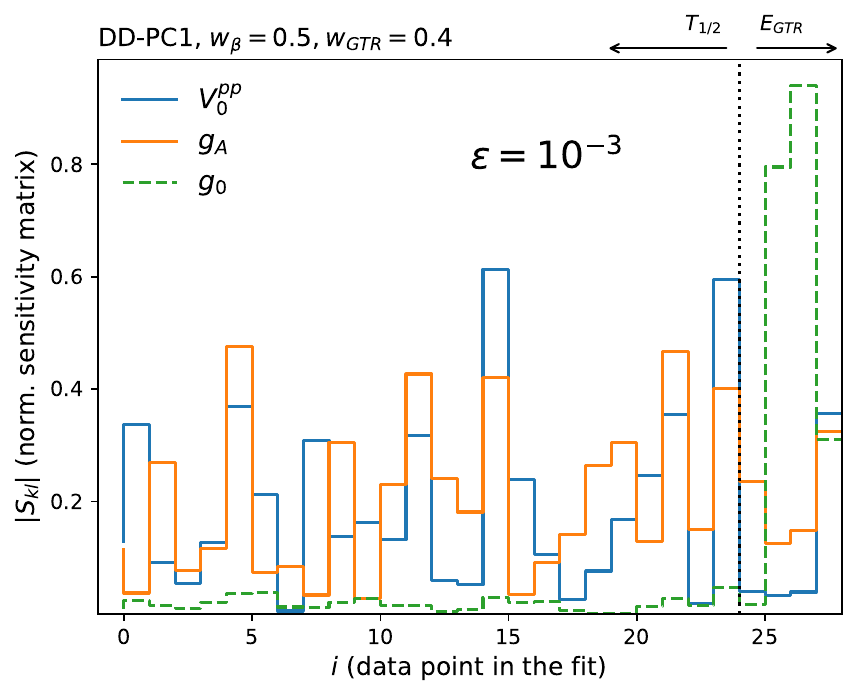}
    \caption{The matrix elements of the normalized sensitivity matrix $|\hat{S}_{ij}|$ for each of the data points included in the optimization. For $i = 0 - 26$ we show the $\beta^-$-decay half-lives and for $i = 27-30$ -- the GT resonance energies.}\label{fig:sensitivity_matrix}
\end{figure}

The covariance matrix is now calculated as
\begin{equation}\label{eq:covariance}
\hat\Sigma(\bar{\boldsymbol{x}}) \approx \chi^2(\bar{\boldsymbol{x}}) \left[\hat{J}^T(\bar{\boldsymbol{x}}) \hat{J}(\bar{\boldsymbol{x}}) \right]^{-1},
\end{equation}
and the parameters' standard deviations are calculated from the diagonal matrix elements of the covariance matrix as $\sigma_k = \sqrt{\hat{\Sigma}(\bar{\boldsymbol{x}})_{k k}}$:
\begin{equation}
    \sigma_{V_0^{pp}} = 0.253, \quad \sigma_{g_A} = 0.347, \quad \sigma_{g_0} = 0.074.
\end{equation}
The correlation matrix 
\begin{equation}
    R_{ij} = \frac{\hat\Sigma(\bar{\boldsymbol{x}})_{ij}}{\sigma_i 
    \sigma_j},
\end{equation}
is equal to:
\begin{equation}
    \hat{R} = \begin{pmatrix}
  1.000    &     -0.737   & 0.041 \\
 -0.737    & 1.000    &        0.157 \\
  0.041  &  0.157  &  1.000       \\
    \end{pmatrix}.
\end{equation}
As expected  \cite{Mustonen2016a},  an anti-correlation between $V_0^{pp}$ and $g_A$ is seen. On the other hand, there is no clear correlation between $V_0^{pp}$ and $g_0$, and between $g_A$ and $g_0$.
Next, we  calculate the sensitivity matrix, defined through the normalized Jacobian $\hat{\tilde{J}}_{kl}(\boldsymbol{y}) = \sigma_l J_{kl}(\boldsymbol{x})$, where $y_k = x_k/\sigma_k$:
\begin{equation}
\hat{S} = \left[ \hat{\tilde{J}}^T(\bar{\boldsymbol{y}}) \hat{\tilde{J}}(\bar{\boldsymbol{y}}) \right]^{-1} \hat{\tilde{J}}^T(\bar{\boldsymbol{y}}).
\end{equation}
The normalized sensitivity matrix is displayed in Fig. \ref{fig:sensitivity_matrix}. It is seen that the $\beta$-decay half-lives are especially sensitive to $V_0^{pp}$ and $g_A$, while the GT resonance centroids are influenced by $g_0$.

 Having determined the optimal parameters $\bar{\boldsymbol{x}} = \begin{pmatrix}
\bar{V}_0^{pp} & \bar{g}_A & \bar{g}_0    
\end{pmatrix}$, we can linearize the half-life around $\bar{\boldsymbol{x}}$ \cite{Dobaczewski2014a}
\begin{equation}
    T_{1/2}^j(\boldsymbol{x}) \approx T_{1/2}^j(\bar{\boldsymbol{x}}) + \boldsymbol{G}^{j} (\boldsymbol{x} - \bar{\boldsymbol{x}}),
\end{equation}
where $j \in \{1^+, 0^-, 1^-, 2^-\}$, corresponding to different transitions, and $\boldsymbol{G}^j = \begin{pmatrix}
    \partial_{V_0^{pp}} T_{1/2} |_{\boldsymbol{p}_0} & \partial_{g_0} T_{1/2} |_{\boldsymbol{p}_0} & \partial_{g_A} T_{1/2} |_{\boldsymbol{p}_0}
\end{pmatrix}$. In practical calculations we vary only $1^-$ and $1^+$ multipoles since they are the most dominant ones. Assuming the Gaussian distribution of weights for each parameter, the uncertainty is calculated as:
\begin{equation}
    (\Delta T_{1/2}^{j})^2 = [\boldsymbol{G}^j]^T \hat{\Sigma} \boldsymbol{G}^j,
\end{equation}
where $\hat\Sigma$ is the covariance matrix of  Eq. (\ref{eq:covariance}). Calculating statistical uncertainties requires evaluating partial derivatives of half-lives with respect to the parameters.

\section{Weak transition rates calculation details}
The nuclear weak decays involve coupling between nuclear and leptonic degrees of freedom. Unlike neutrinos, whose wave functions can be represented with a plane-wave solution to the Dirac equation, the electron (or positron) wave functions have to be obtained by directly solving the Dirac equation with nuclear charge contribution. We will refer to these functions as electron radial wave functions (ERWFs). In this work, we resort to the formalism introduced in Refs. \cite{Behrens1971a,Bambynek1977a,Behrens1970a} where  the ERWFs are expanded in the powers of electron mass ($m_e$), electron energy multiplied by the nuclear radius ($E_x R$) and nuclear charge ($\alpha Z$), and keep those powers at the leading order (LO). Alternatively, one can also obtain ERWFs directly by solving the Dirac equation \cite{Koshigiri1979a,Horiuchi2021a}. 
The relevant expressions are derived and tabulated in Refs. \cite{Bambynek1977a,Behrens1970a,Behrens1971a}. We have compared the rates for superheavy nuclei using exact ERWFs and the LO approximation and concluded that the associated error  does not exceed  30\% for lifetimes. The details of this implementation and comparison between these two methods will be presented in a forthcoming publication \cite{Ravlic2024b}.


\subsection{Contribution of different operators to the total EC rate}
Taking the $_{108}$Hs isotopic chain as an example, we study the contribution of different operators to the total EC $1^-$ rate. At the LO, a total of 5 operators contribute to the decay rate, which leads to 15 matrix elements including interference terms. Those operators include:
\begin{description}
\item[Spin-dipole operator and  spin-dipole operator with finite-size correction]
\begin{equation}
    \hat{F}(\text{SD}) = \sqrt{2}g_A r [\boldsymbol{\sigma} \otimes \boldsymbol{C}_1]_1, \quad  \hat{F}(\text{FSSD}) = \frac{2\sqrt{2}}{3}g_A I(1,1,1,1;r)r [\boldsymbol{\sigma} \otimes \boldsymbol{C}_1]_1,
\end{equation}
\item[Dipole operator and its finite-size correction]
\begin{equation}
    \hat{F}(\text{DIP}) = r \boldsymbol{C}_1, \quad  \hat{F}(\text{FSDIP}) = \frac{2}{3}I(1,1,1,1;r) \boldsymbol{C}_1,
\end{equation}
\item[Relativistic correction to the rate]
\begin{equation}
    \hat{F}(\text{REL}) = \boldsymbol{\alpha}.
\end{equation}
\end{description}

In the above expressions, $I(1,1,1,1;r)$ is the finite-size correction radial function defined in Ref. \cite{Bambynek1977a}, $C_{LM} =\sqrt{4\pi /(2L+1)}Y_{LM}$ is the normalized spherical harmonic, and $[\boldsymbol{\sigma} \otimes \boldsymbol{C}_1]_1$ denotes coupling of Pauli spin matrix and normalized spherical harmonic to a tensor of rank 1.

\begin{figure}
    \centering
    \includegraphics[width=0.5\linewidth]{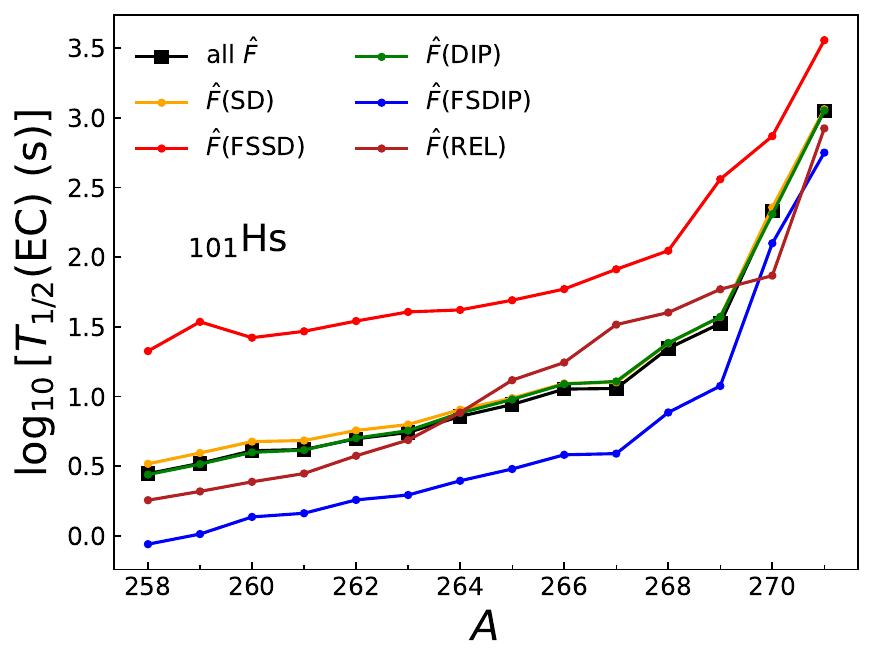}
    \caption{The $1^-$ contribution to the EC half-life in ${}_{108}$Hs. The total result, where the contribution of all $1^-$ operators is included (squares), is compared to the case where an operator is excluded from the half-life calculation.}
    \label{fig:1m_contribution}
\end{figure}

In order to test which of these operators contributes the most to the half-life of $1^-$ multipole, we calculate  half-lives  by removing one-by-one  the operators from the matrix element.
The results are displayed in  Fig.~\ref{fig:1m_contribution} for the Hs chain.
(We note that removing a specific operator  removes all the interference terms corresponding to that operator.) First, the removal of spin-dipole (SD) and dipole (DIP) operators from the calculation has almost no impact on  half-lives. Removing the relativistic $\boldsymbol{\alpha}$ operator (REL) has some effect. The most significant correction stems from the operators containing the finite-size correction $I(1,1,1,1;r)$. In fact, removing the finite-size spin-dipole (FSSD) operator increases the half-lives by almost an order of magnitude, while removing the finite-size dipole (FSDIP) operator further decreases the half-life. Both an increase and decrease of half-lives are possible since these terms can interfere coherently or destructively for a particular excitation. Since the SD operators originate from axial-vector current and DIP operators from vector current, we observe that both are contributing to the weak decay rate.

\subsection{$\pmb{\beta^\pm}$-decay calculation caveats}\label{subsec:beta_caveats}

\begin{figure}
    \centering
    \includegraphics[width=0.9\linewidth]{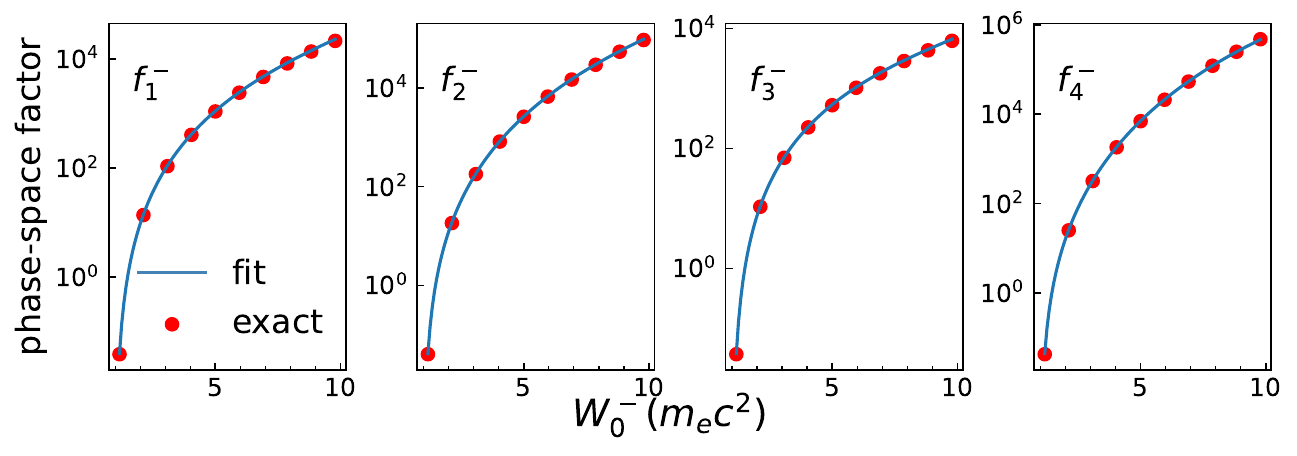}
    \caption{The $\beta^-$-decay phase-space factors $f_{1-4}(W_0^-)$ calculated exactly as in Eq. (\ref{eq:phase_space_factor}) (circles), or by Lagrange interpolation on a Chebyshev grid (line). Calculations are performed for ${}^{286}$Rf. }
    \label{fig:phase-space-factor}
\end{figure}

 Unlike the EC, where an electron is in a discrete orbital $x$, in the case of $\beta^\pm$-decay the electron is embedded in the continuum, and the summation over $x$ is replaced by integration over electron energy $W$. The contour integral takes the form:
\begin{equation}
    \lambda_{\beta^\pm} = \frac{\text{ln}2}{K} \frac{1}{2\pi i} \oint  d \omega \int \limits_1^{W_0^\pm[\omega]} dW p W (W_0^\pm[\omega] - W)^2 F(\mp |Z|, W) C(W, \omega),
\end{equation}
where the end-point energy $W_0^\pm[\omega]$ is
\begin{equation}
    W_0^+[\omega](m_ec^2) = \omega -\lambda_p + \lambda_n - \Delta_{nH} - 2m_ec^2, \quad W_0^-[\omega](m_e c^2) = \omega - \lambda_n + \lambda_p + \Delta_{np},
\end{equation}
where $\Delta_{np} = 1.293$ MeV is the neutron-proton mass difference, $\lambda_{n(p)}$ is the neutron(proton) chemical potential, $\Delta_{nH} = 0.782$ MeV is the neutron-hydrogen atom mass difference, and $m_e$ is the electron mass. The shape-factor has the usual form \cite{Behrens1970a,Behrens1971a,Bambynek1977a}
\begin{equation}
    C(W, \omega) = k(\omega) + ka(\omega) W + \frac{kb(\omega)}{W} + kc(\omega) W^2,
\end{equation}
with $k , ka, kb$, and $kc$ defined in \cite{Behrens1970a,Behrens1971a,Bambynek1977a}. We can write the contour integral  as:
\begin{equation}\label{eq:rate_with_phase_space}
    \lambda_{\beta^\pm} = \frac{\text{ln}2}{K} \frac{1}{2\pi i} \oint  d \omega \left\{  f^\pm_1(W_0^\pm[\omega]) k(\omega) +  f^\pm_2(W_0^\pm[\omega]) ka(\omega) +  f^\pm_3(W_0^\pm[\omega]) kb(\omega) +  f^\pm_4(W_0^\pm[\omega]) kc(\omega)  \right\},
\end{equation}
where the phase-space factors are
\begin{align}\label{eq:phase_space_factor}
    \begin{split}
        &f_1^\pm(W_0^\pm[\omega]) = \int \limits_{1}^{W_0^\pm[\omega]} dW pW(W_0^\pm[\omega] - W)^2 F(\mp |Z|, W), \quad f^\pm_2(W_0^\pm[\omega]) = \int \limits_{1}^{W_0^\pm[\omega]} dW pW^2(W_0^\pm[\omega] - W)^2 F(\mp |Z|, W), \\
        &f^\pm_3(W_0^\pm[\omega]) = \int \limits_{1}^{W_0^\pm[\omega]} dW p(W_0^\pm[\omega] - W)^2 F(\mp |Z|, W), \quad f^\pm_4(W_0^\pm[\omega]) = \int \limits_{1}^{W_0^\pm[\omega]} dW pW^3(W_0^\pm[\omega] - W)^2 F(\mp |Z|, W), \\
    \end{split}
\end{align}
differing only by the  power of $W$. The problem in applying the contour integration directly as in Eq. (\ref{eq:rate_with_phase_space}) is that $f_{1-4}^\pm$ are non-analytic due to properties of Fermi function $F(Z,W)$. Therefore, to perform the integration, we follow the strategy advocated in Refs. \cite{Mustonen2016a,Ney2020a} where the phase-space factor is interpolated using Lagrange  polynomials on a Chebyshev grid, which can subsequently be integrated using simple quadrature rules. This strategy is illustrated in Fig. \ref{fig:phase-space-factor}, which shows the phase-space factors for $\beta^-$ calculated directly using Eq. (\ref{eq:phase_space_factor}) and the Lagrange interpolation. Note that although $f_{1-4}^-$ span several orders of magnitude, the interpolation performs very well.

\newpage\clearpage

\section{Table of EC/$\pmb{\beta^+}$ branching ratios}

\begin{table}[htb]
\caption{The list of nuclei in the region $101 \leq Z \leq 118$ with the ratio between experimentally measured and calculated  half-lives is larger than 0.05, i.e.,  $\mathcal{R} = T_{1/2}^{exp}/T_{1/2}^{calc} > 0.05$. The corresponding standard deviations $\Delta \mathcal{R}$ are also shown.}\label{betam-table}
    \begin{tabular}{ccc|ccc}
    \hline
    \hline
    $Z$ & nucleus & $\mathcal{R} \pm \Delta \mathcal{R}$ & $Z$ & nucleus & $\mathcal{R} \pm \Delta \mathcal{R}$ \\
    \hline
101 & ${}^{244}$Md & $ 0.05 \pm 0.02 $ & 101 & ${}^{246}$Md & $ 0.05 \pm 0.02 $ \\
101 & ${}^{247}$Md & $ 0.06 \pm 0.02 $ & 101 & ${}^{248}$Md & $ 0.22 \pm 0.12 $ \\
101 & ${}^{249}$Md & $ 0.57 \pm 0.17 $ & 101 & ${}^{250}$Md & $ 0.53 \pm 0.15 $ \\
101 & ${}^{251}$Md & $ 2.03 \pm 0.51 $ & 101 & ${}^{252}$Md & $ 0.41 \pm 0.19 $ \\
101 & ${}^{253}$Md & $ 0.7 \pm 0.51 $ & 102 & ${}^{253}$No & $ 1.52 \pm 0.39 $ \\
102 & ${}^{254}$No & $ 0.16 \pm 0.05 $ & 102 & ${}^{255}$No & $ 0.39 \pm 0.11 $ \\
103 & ${}^{254}$Lr & $ 0.32 \pm 0.12 $ & 103 & ${}^{255}$Lr & $ 0.36 \pm 0.16 $ \\
103 & ${}^{256}$Lr & $ 0.22 \pm 0.13 $ & 103 & ${}^{260}$Lr & $ 0.17 \pm 0.06 $ \\
104 & ${}^{255}$Rf & $ 0.11 \pm 0.04 $ & 104 & ${}^{257}$Rf & $ 0.13 \pm 0.05 $ \\
104 & ${}^{263}$Rf & $ 0.10 \pm 0.05 $ & 105 & ${}^{256}$Db & $ 0.37 \pm 0.17 $ \\
105 & ${}^{257}$Db & $ 0.32 \pm 0.14 $ & 105 & ${}^{258}$Db & $ 0.15 \pm 0.08 $ \\
105 & ${}^{260}$Db & $ 0.06 \pm 0.02 $ & 105 & ${}^{261}$Db & $ 0.13 \pm 0.05 $ \\
105 & ${}^{262}$Db & $ 0.74 \pm 0.35 $ & 105 & ${}^{263}$Db & $ 0.30 \pm 0.11 $ \\
105 & ${}^{266}$Db & $ 3.94 \pm 3.61 $ & 106 & ${}^{259}$Sg & $ 0.08 \pm 0.03 $ \\
106 & ${}^{263}$Sg & $ 0.05 \pm 0.02 $ & 106 & ${}^{265}$Sg & $ 0.23 \pm 0.08 $ \\
106 & ${}^{267}$Sg & $ 0.30 \pm 0.15 $ & 107 & ${}^{264}$Bh & $ 0.13 \pm 0.05 $ \\
107 & ${}^{265}$Bh & $ 0.10 \pm 0.05 $ & 107 & ${}^{266}$Bh & $ 0.77 \pm 0.30 $ \\
107 & ${}^{267}$Bh & $ 0.65 \pm 0.34 $ & 108 & ${}^{264}$Hs & $ 0.12 \pm 0.07 $ \\
108 & ${}^{268}$Hs & $ 0.08 \pm 0.06 $ & 108 & ${}^{269}$Hs & $ 0.57 \pm 0.36 $ \\
108 & ${}^{270}$Hs & $ 0.06 \pm 0.04 $ & 109 & ${}^{270}$Mt & $ 0.07 \pm 0.05 $ \\
111 & ${}^{282}$Rg & $ 0.37 \pm 0.19 $ & 113 & ${}^{286}$Nh & $ 0.06 \pm 0.04 $ \\
114 & ${}^{290}$Fl & $ 0.22 \pm 0.21 $ & 115 & ${}^{290}$Mc & $ 0.12 \pm 0.07 $ \\
        \hline
\end{tabular}
\end{table}

\section{Other $\pmb{\beta^-}$-decay chains in superheavy region}
\begin{figure}[h!]
    \includegraphics[width=0.85\linewidth]{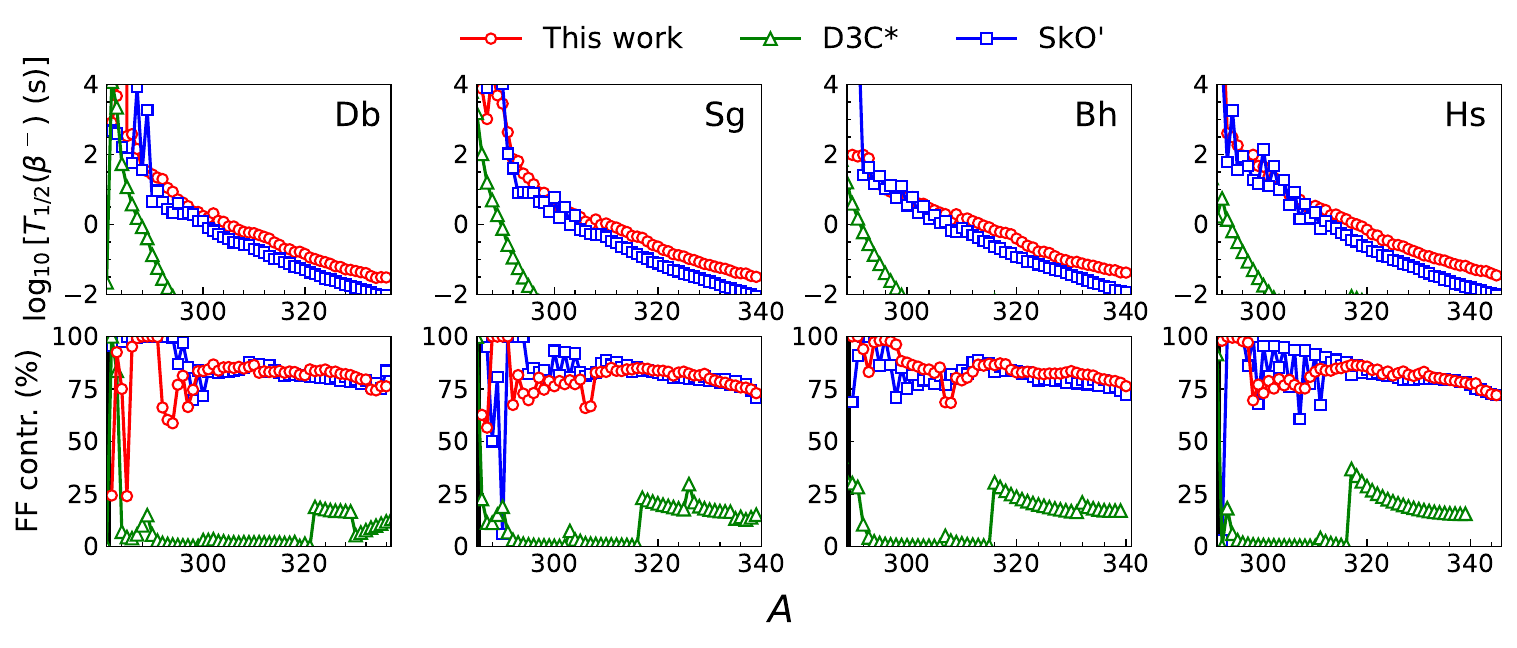}
    \caption{(Top) The $\beta^-$-decay half-lives of selected superheavy isotopic chains. (Bottom) Contribution of first-forbidden transitions to the total decay rate. 
    Results obtained in this paper are  compared to those from Refs. \cite{Ney2020a,Marketin2016a}.}
    \label{fig:other_chains}
\end{figure}

Figure \ref{fig:other_chains}  shows the results for the $\beta^-$-decay half-lives of 
Db, Sg, Bh, and Hs chains compared  to the results of Refs. \cite{Ney2020a,Marketin2016a}. Upper panels show the half-lives, and lower panels the contribution of first-forbidden transitions (in \%).

\newpage

%